\documentclass[journal=jacsat, manuscript=article, layout=traditional]{achemso}

\usepackage[version=3]{mhchem} 
\usepackage{plimsoll} 

\def\lmo{\ce{LaMnO3(001)}}
\def\pt{\ce{Pt(111)}}
\def\ushe{U_{\text{SHE}}}
\def\textpH{\text{pH}}
\def\grandpot{\Omega_{\text{surf}}(U_{\text{SHE}}, \text{pH})}
\def\elemA{\mathrm{A}}
\def\kb{k_{\text{B}}}
\def\muA{\mu_\elemA}
\def\muspecies{\mu_\mathrm{H_x AO_y^{z-}}}
\def\logconc{\log_{10} c_\mathrm{H_x A O_y^{z-}}}
\def\conc{c_\mathrm{H_x A O_y^{z-}}}

\author{Xiaochen Du}
\affiliation[ChemE]{Department of Chemical Engineering, Massachusetts Institute of Technology, Cambridge, MA 02139, USA}
\author{Mengren Liu}
\author{Jiayu Peng}
\author{Hoje Chun}
\author{Alexander Hoffman}
\author{Bilge Yildiz}
\affiliation[DMSE]{Department of Materials Science and Engineering, Massachusetts Institute of Technology, Cambridge, MA 02319, USA}
\alsoaffiliation{Department of Nuclear Science and Engineering, Massachusetts Institute of Technology, Cambridge, MA 02319, USA}
\author{Lin Li}
\affiliation[LL]{Massachusetts Institute of Technology Lincoln Laboratory, Lexington, MA, 02421, USA}
\author{Martin Z. Bazant}
\affiliation[ChemE]{Department of Chemical Engineering, Massachusetts Institute of Technology, Cambridge, MA 02139, USA}
\author{Rafael Gómez-Bombarelli}
\email{rafagb@mit.edu}
\affiliation[DMSE]{Department of Materials Science and Engineering, Massachusetts Institute of Technology, Cambridge, MA 02319, USA}

\title{Accelerating and enhancing thermodynamic simulations of electrochemical interfaces}

\begin{document}


\begin{abstract}
  Electrochemical interfaces are crucial in catalysis, energy storage, and corrosion, where their stability and reactivity depend on complex interactions between the electrode, adsorbates, and electrolyte. Predicting stable surface structures remains challenging, as traditional surface Pourbaix diagrams tend to either rely on expert knowledge or costly \textit{ab initio} sampling, and neglect thermodynamic equilibration with the environment. Machine learning (ML) potentials can accelerate static modeling but often overlook dynamic surface transformations. Here, we extend the Virtual Surface Site Relaxation-Monte Carlo (VSSR-MC) method to autonomously sample surface reconstructions modeled under aqueous electrochemical conditions. Through fine-tuning foundational ML force fields, we accurately and efficiently predict surface energetics, recovering known \pt{} phases and revealing new \lmo{} surface reconstructions. By explicitly accounting for bulk-electrolyte equilibria, our framework enhances electrochemical stability predictions, offering a scalable approach to understanding and designing materials for electrochemical applications.
\end{abstract}


\section{Introduction}
The structure of surfaces and interfaces critically influences the performance and stability of materials in applications ranging from (electro-)catalysis \cite{xie_surface_2020, zhang_hydrogen-induced_2022, sumaria_atomic-scale_2023} and energy storage \cite{xu_energy_2016, xu_bulk_2021, xiao_understanding_2020} to electronic devices \cite{monch2013semiconductor, luth2013surfaces}. In gas or vacuum environments, a nominally pristine surface derived by cleaving the material bulk often undergoes atomic rearrangements that can produce intricate reconstructions, as documented for materials such as Si \cite{binnig_7_1983, csanyi_learn_2004, shen_deciphering_2023}, GaN \cite{northrup_structure_2000, kusaba_exploration_2022}, and $\alpha$-\ce{Al2O3} \cite{hutner_stoichiometric_2024}. Under aqueous electrochemical conditions, other factors influence the stability of the surface: the electrolyte pH, the applied electrical potential, the presence of adsorbates such as OH* and \ce{H2O}*, \cite{may_influence_2012, seitz_highly_2016, fabbri_dynamic_2017, song_dissolution-induced_2019, zhang_dissolutionprecipitation_2019, wan_amorphization_2021, weber_atomistic_2022, peng_stability_2022, peng_design_2023, lu_key_2024} and the formation of charged layers \cite{hormann_grand_2019, mathew_implicit_2019, ringe_implicit_2022}. These added variables drive the dissolution and re-deposition of metal surface atoms, posing significant challenges for understanding and predicting the structure and stability of electrochemical interfaces \cite{auer_self-activation_2020, stevens_new_2022, peng_toward_2025}.

Pourbaix diagrams, which delineate thermodynamically-favored phases as a function of pH and electrical potential, are a cornerstone in assessing electrochemical material stability. An individual species Pourbaix diagram shows the most stable bulk or dissolved species of an element, whereas a surface Pourbaix diagram illustrates the most stable surface structure given a fixed bulk phase. Both types of Pourbaix diagrams have guided the design of corrosion prevention strategies \cite{delahay_potential-ph_1950, huang_reliable_2019}, informed the development of electrocatalysts \cite{hansen_surface_2008, chatenet_water_2022, du_interface_2022, chen_reconstructed_2023, zhao_formation_2024, wu_pivotal_2024, peng_toward_2025}, and aided in planning rational synthesis routes \cite{sun_non-equilibrium_2019, wang_optimal_2024}. 

In recent decades, first-principles computational methods have enhanced the ability to construct surface Pourbaix diagrams by predicting atomic-level interfacial energetics \cite{hansen_surface_2008, zeng_towards_2015, vinogradova_quantifying_2018, sun_non-equilibrium_2019}. However, many such studies were limited to pristine facets with pre-defined adsorbate coverages, neglecting more complex reconstructions or adsorption patterns under operational conditions \cite{greeley_computational_2006, hansen_surface_2008, bajdich_theoretical_2013, williams_first_2014, vinogradova_quantifying_2018, hao_lanthanide-doped_2023, morankar_first_2023}. Moreover, traditional Pourbaix diagrams often assume a fixed concentration for dissolved species (commonly $10^{-5}$ to $10^{-6}$ M) \cite{persson_prediction_2012, rong_ab_2015, singh_electrochemical_2017, singh_robust_2019, sun_non-equilibrium_2019, wang_predicting_2020, wang_optimal_2024, lee_machine-learning-accelerated_2025}. While computationally convenient, this assumption disregards true equilibration between the bulk phase and its dissolved ions. As a result, important stability behaviors may be obscured, especially when designing long-lifetime catalysts based on bulk-electrolyte equilibration principles \cite{kan_accelerated_2024, jjenewein_automated_2024} or synthesizing materials that are only stable at higher dissolved-ion concentrations \cite{wang_optimal_2024}.

Studies considering surface dissolution, re-deposition, or interchanges within multi-component systems often rely on “hand-crafted” structural models validated by costly first-principles calculations, a method that becomes insufficient for exploring complex multi-component systems with large compositional and configurational spaces \cite{rong_ab_2015, seitz_highly_2016, qiu_ab_2018, lu_key_2024}. Recent efforts to sample across the compositions of a binary system nevertheless relied on costly \textit{ab initio} grand canonical Monte Carlo (GCMC) sampling that might not scale to more complex materials \cite{qin_unveiling_2024}.

Machine learning (ML) techniques, especially neural network force fields (NFFs), offer a pathway to address these challenges by enabling faster atomistic modeling \cite{ulissi_machine-learning_2017, back_convolutional_2019, flores_active_2020, unke_machine_2021, axelrod_learning_2022, kocer_neural_2022, noordhoek_accelerating_2024, tran_rational_2024}, but current approaches overlook complex surface reconstruction phenomena. The growing availability of pre-trained, foundational NFFs expands the applicability of these methods to new material classes with limited density-functional theory (DFT) data \cite{chen_universal_2022, deng_chgnet_2023, batatia_foundation_2024, neumann_orb_2024, shoghi_molecules_2024, barroso-luque_open_2024}. However, most ML-based approaches in catalysis either focus on high-throughput screening of ideal surfaces with pre-defined adsorbate patterns \cite{tran_active_2018, back_toward_2019, chanussot_open_2021, sanspeur_wherewulff_2023, tran_open_2023, tran_rational_2024, lunger_towards_2024} or examine reconstruction only within a narrow set of configurations---permitting select adsorbate and metal surface combinations but disallowing general surface dissolution and re-deposition---thus leaving the broader compositional space largely unexplored \cite{ulissi_automated_2016, ghanekar_adsorbate_2022, bang_machine_2023, sharma_machine-learning-assisted_2024, lee_machine-learning-accelerated_2025, zheng_active_2025}.

Adapting thermodynamic sampling approaches developed for gas- or vacuum-exposed surfaces, such as the Virtual Surface Site Relaxation-Monte Carlo (VSSR-MC) algorithm \cite{du_machine-learning-accelerated_2023}, can capture more realistic reconstruction phenomena. Extending such approaches to electrochemical conditions requires accounting for pH, applied electrical potential, and relevant dissolved species concentrations, along with a robust treatment of the complex equilibria between the electrolyte, the electrode surface, and the electrode bulk.

In this work, we develop a comprehensive methodology that addresses these challenges in three ways. First, we adapt the VSSR-MC sampling approach to electrochemical conditions, enabling automated discovery of stable surface reconstructions across compositional and configurational spaces involving both adsorbates and underlying metal species. Second, we demonstrate how pre-trained NFFs can be fine-tuned to achieve DFT-level accuracy for surface energetics while maintaining computational efficiency. Third, we establish a theoretical framework for calculating surface Pourbaix diagrams that explicitly accounts for dynamic equilibria with the electrode bulk and aqueous species at varying concentrations. 

We validate our approach through two case studies of increasing complexity. Using \pt{} as a benchmark, we demonstrate the ability to efficiently sample known surface reconstructions while confirming the accuracy of the fine-tuned NFF. We then tackle the more challenging case of \lmo{} and uncover reconstructions absent from prior work. Finally, we highlight the importance of capturing electrolyte-bulk equilibria by constructing surface Pourbaix diagrams that respect the thermodynamic coupling between electrolyte species, the electrode surface, and the electrode bulk. These case studies showcase the ability of our method to predict realistic electrochemical interfaces, accelerating the discovery of stable surfaces for energy conversion and catalysis.

\section{Results}
\begin{figure}[ht!]
  \centering
  \includegraphics[width=1.0\textwidth]{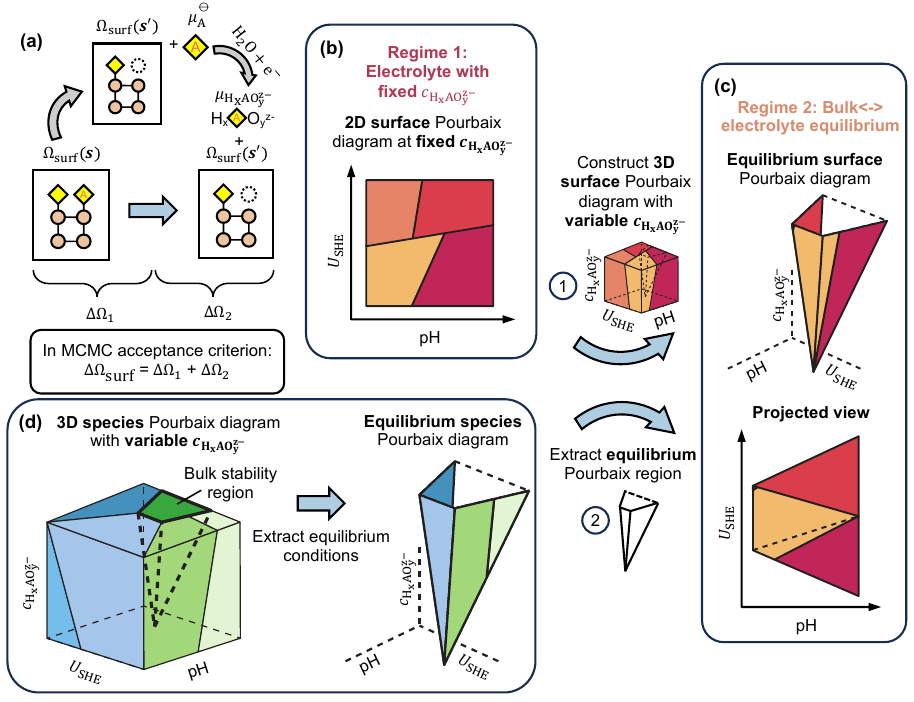}
  \caption{\textbf{Thermodynamic analysis for electrochemical interfaces.} (a) Two-step process for comparing surface Pourbaix grand potential differences ($\Delta \grandpot$) between two slabs. The figure shows the case of a single-atom dissolution, but more complicated changes can be constructed as a sequence of dissolution/adsorption steps. (b-c) Two regimes for constructing surface Pourbaix diagrams. (b) Conventional surface Pourbaix diagram at fixed dissolved species concentrations. (c) Equilibrium surface Pourbaix diagram derived from bulk-electrolyte equilibrium conditions. (d) To construct the equilibrium surface Pourbaix diagram, the 3D species Pourbaix diagram is generated to define the equilibrium conditions for bulk stability in pH-$\ushe$-$\conc$ space. Afterwards, a 3D surface Pourbaix diagram is constructed from the surface Pourbaix diagram in (b) and the equilibrium Pourbaix region traced out to produce the final equilibrium surface Pourbaix diagram.}
\label{fig:method}
\end{figure}

\subsection{Sampling and evaluating aqueous electrochemical interfaces}
We modified VSSR-MC \cite{du_machine-learning-accelerated_2023} to sample multi-atom adsorbates such as \ce{OH}* and \ce{H2O}*, in addition to single-atom adsorbates such as O* and metal species, at virtual sites (see Methods ``Surface slab modeling''). Additionally, we included the pristine surface atoms as virtual sites (see Figs.~S2(b) and~S5(b)) to simulate dissolution and re-deposition, thereby broadening the search space of possible reconstructions.

In each Monte Carlo (MC) step, the energy difference between the proposed and current slab is calculated via a two-step process to account for the dominant species of each element under aqueous electrochemical conditions \cite{rong_ab_2015, qiu_ab_2018}. In Fig.~\ref{fig:method}(a), we consider the case of a single atom dissolution; for more complex MC steps or to compare stability between two arbitrary surfaces, a sequence of dissolution/adsorption steps is constructed to calculate the total energy change. In step 1, atom A is removed from the surface---creating a vacancy---and released in its standard state, yielding energy change $\Delta \Omega_1$. In step 2, the freed atom converts to the most stable aqueous species \ce{H_x AO_y^{z-}}, following the computational hydrogen electrode (CHE) framework \cite{norskov_origin_2004}, with energy change $\Delta \Omega_2$. The overall energy difference is given by:
\begin{align}
    \Delta \Omega_\text{surf} (\mathbf{X}, \{\mu\}, T, \ushe, \textpH) = \Delta \Omega_1 + \Delta \Omega_2
\end{align}
where $\mathbf{X}$ refers to the atomic identities and positions, $\{\mu\}$ represents the set of chemical potentials 
of the stable aqueous species $\mu_{\mathrm{H_x A O_y^{z-}}}$ (related to species concentration), $T$ is the temperature, and $\ushe$ is the electrical potential relative to the standard hydrogen electrode (SHE). The surface Pourbaix grand potential, $\grandpot$, of each slab can thus be interpreted as the surface formation energy derived from the stable aqueous species at a given $\ushe$ and pH (with fixed aqueous species concentrations and temperature). See ``Surface Pourbaix grand potential'' in Methods for details.

We constructed surface Pourbaix diagrams under two thermodynamic regimes. In regime 1 (Fig.~\ref{fig:method}(b)), the conventional surface Pourbaix diagram is plotted at a fixed dissolved species concentration ($\conc$) in the electrolyte, which acts as the thermodynamic reservoir. Here, stable aqueous species for each metal atom are identified at each pH and $\ushe$ using the species Pourbaix diagram. Since each domain in the species Pourbaix diagram corresponds to a distinct set of dominant species, we compute separate $\grandpot$ values for all considered surface slabs within each domain. A convex hull analysis of these grand potentials reveals the stable surface domains \cite{persson_prediction_2012, singh_electrochemical_2017, patel_efficient_2019, wang_predicting_2020}, which may be the same across species domain boundaries. Thus, all surface domains are pooled and merged to produce the final surface Pourbaix diagram. A detailed schematic is provided in Fig.~S1.

In regime 2 (Fig.~\ref{fig:method}(c)), we perform a bulk-electrolyte equilibrium analysis where both the electrode bulk and the electrolyte serve as thermodynamic reservoirs in equilibrium with the surface. In this regime, we plot the equilibrium surface Pourbaix diagram by extending the conventional Pourbaix diagram into three dimensions to explicitly incorporate $\conc$ as a variable dependent on pH and $\ushe$. First, a three-dimensional (3D) species Pourbaix diagram is generated with an additional, independent $\conc$ axis. By following the phase boundaries of the bulk stability region in pH-$\ushe$-$\conc$ space, we trace out the equilibrium species Pourbaix diagram that captures the range of conditions under which the electrode bulk remains stable and in equilibrium with the neighboring electrolyte species. In this equilibrium Pourbaix diagram, pH, $\ushe$, and $\conc$ are thermodynamically coupled rather than independent variables. See Fig.~\ref{fig:method}(d) for an illustration and ``Thermodynamic equilibria'' in Methods for details. The equilibrium surface Pourbaix diagram is constructed analogously by generating a 3D surface Pourbaix diagram (using a workflow similar to Fig.~S1) and delineating the same bulk stability boundaries (details also in Methods ``Thermodynamic equilibria''). Projecting the 3D equilibrium surface Pourbaix diagram into the pH-$\ushe$ plane yields a diagram that resembles a conventional Pourbaix diagram but with $\conc$ as a dependent variable.

\begin{figure}[ht!]
  \centering
  \includegraphics[width=1.0\textwidth]{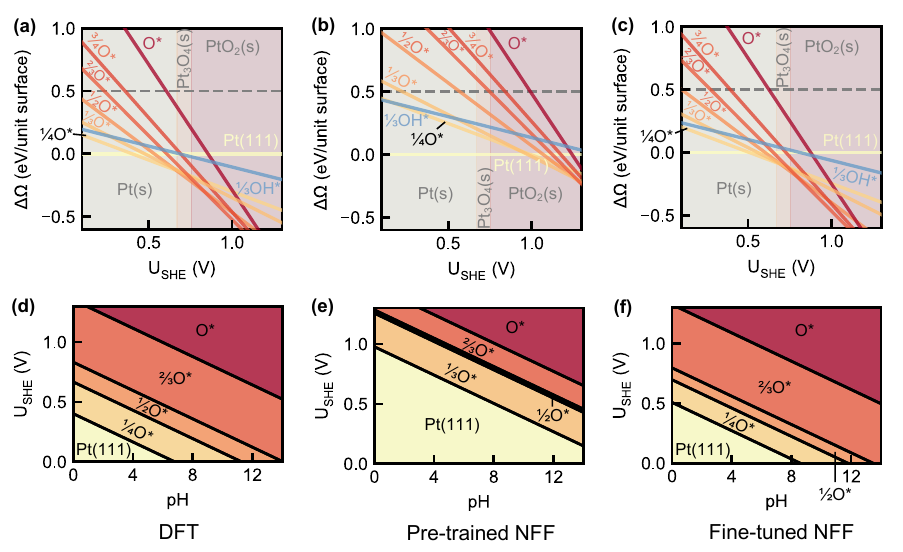}
  \caption{\textbf{\pt{} energy analysis and surface Pourbaix diagrams with various energy models.} (a-c) $\Delta \grandpot$ comparison of handpicked Pt(111) surfaces with respect to the pristine surface in eV/surface unit cell across $\ushe$ at fixed pH = 0 with (a) DFT, (b) pre-trained NFF (CHGNet), and (c) fine-tuned NFF (MACE) energies. The bulk species domains are differentiated through the background color and labeled in gray. Dotted lines are a guide for the eye. (d-f) Pt(111) surface Pourbaix diagrams for (d) DFT, (e) pre-trained NFF, and (f) fine-tuned NFF.}
\label{fig:pt}
\end{figure}

\subsection{Validating modified VSSR-MC and fine-tuning}
\subsubsection{\pt}
Pt is a well-known catalyst used in thermo- and electro-catalysis \cite{sumaria_atomic-scale_2023,xu_atomistic_2022}. We selected 8 handpicked Pt surface structures, including the pristine surface, from literature \cite{hansen_surface_2008, vinogradova_quantifying_2018}, featuring a mix of O* and stoichiometric \ce{[OH-H2O]}* adsorption levels across three supercell sizes: $2\times 2$, $3\times 3$, and $\sqrt{3}\times 3$ (see Fig.~S2(a-b) for the pristine surfaces and Fig.~S2(c) for specific structures). We denote the adlayer of a given adsorbate with the fractional coverage ($\theta$, the number of adsorbates per exposed metal atom) preceding the adsorbate of interest (e.g., $\frac{1}{4}$O*).

We used CHGNet \cite{deng_chgnet_2023} as our pre-trained NFF model, as its training set already incorporates energy corrections from the Materials Project 2020 compatibility scheme \cite{wang_framework_2021}. This choice ensures seamless integration with the Materials Project aqueous compatibility scheme, which we used to align energies for \(\grandpot\) \cite{persson_prediction_2012, singh_electrochemical_2017, patel_efficient_2019}. (See “Species Pourbaix diagram construction” and “Surface Pourbaix grand potential” in Methods.) With pre-trained CHGNet, the relative surface Pourbaix grand potential ($\Delta \grandpot$) of these O* and \ce{[OH-H2O]}* levels with respect to the pristine surface were obtained at fixed pH = 0 and varying $\ushe$ (Fig.~\ref{fig:pt}(b)). These grand potentials correctly increase with increasing O* levels but the values are approximately twice as high as the DFT reference (Fig.~\ref{fig:pt}(a)). CHGNet overpredicts these grand potentials because it was pre-trained on the MPtrj dataset---consisting only of bulk crystals\cite{deng_materials_2023}---and our Pt surface system with O* and OH* adsorbates is thus out-of-distribution. The adsorbate phases for the pre-trained CHGNet surface Pourbaix diagram (at $10^{-6}$ M dissolved species concentration in Fig.~\ref{fig:pt}(e)) are shifted towards higher $\ushe$ and pH compared to the DFT reference in Fig.~\ref{fig:pt}(d). Moreover, pre-trained CHGNet does not predict the correct adlayer phases (e.g., $\frac{1}{3}$O* instead of $\frac{1}{4}$O*) since the relative $\Delta \grandpot$ among the handpicked surfaces are not sufficiently precise. 

We performed VSSR-MC sampling in all three slab sizes starting from the pristine surfaces with algorithmically-generated virtual sites \cite{du_machine-learning-accelerated_2023} (see ``VSSR-MC'' in Methods for details) allowing for the adsorption and desorption of Pt, O, OH, and \ce{H2O}. An example sampling profile can be found in Fig.~S3. When the pre-trained CHGNet \cite{deng_chgnet_2023} was used as the surrogate model, we obtained all handpicked surfaces from literature with this sampling approach. 

For fine-tuning, we downsampled 115 sampled structures and additionally rattled the dominant structures found in the pre-trained CHGNet surface Pourbaix diagram (Fig.~\ref{fig:pt}(e)) for another 40 structures to obtain a total of 155 structures. We fine-tuned both CHGNet and MACE \cite{batatia_mace_2022, batatia_foundation_2024} and found that MACE performed significantly better than CHGNet after fine-tuning to predict surface Pourbaix diagrams, despite being pre-trained on uncorrected DFT energies. Fig.~S4 compares the performance of both fine-tuned models against the pre-trained CHGNet. After fine-tuning, the predicted energy mean-absolute error (MAE) is within 4 meV/atom for the known surfaces and thus much closer to the DFT reference. Fine-tuning details are in Methods ``Active learning''.

With the fine-tuned NFF, predicted $\Delta \grandpot$ were within 50 meV/surface unit cell, where the four-atom thick $1\times1$ unit cell serves as the normalization reference across all slab sizes. This level of accuracy, shown in Fig.~\ref{fig:pt}(c), is sufficient to discern the subtle energetic differences among different phases. Small changes in relative energies can cause deviations in the predicted ordering of surfaces, but our fine-tuned NFF correctly predicts close phase boundaries compared to the DFT reference (Fig.~\ref{fig:pt}(f)). Overall, our analysis of Pourbaix diagrams with fine-tuned MACE on one of the simpler electrocatalytic systems indicates that fine-tuned NFFs can match DFT energy predictions for surface structures.

\begin{figure}[ht!]
  \centering
  \includegraphics[width=1.0\textwidth]{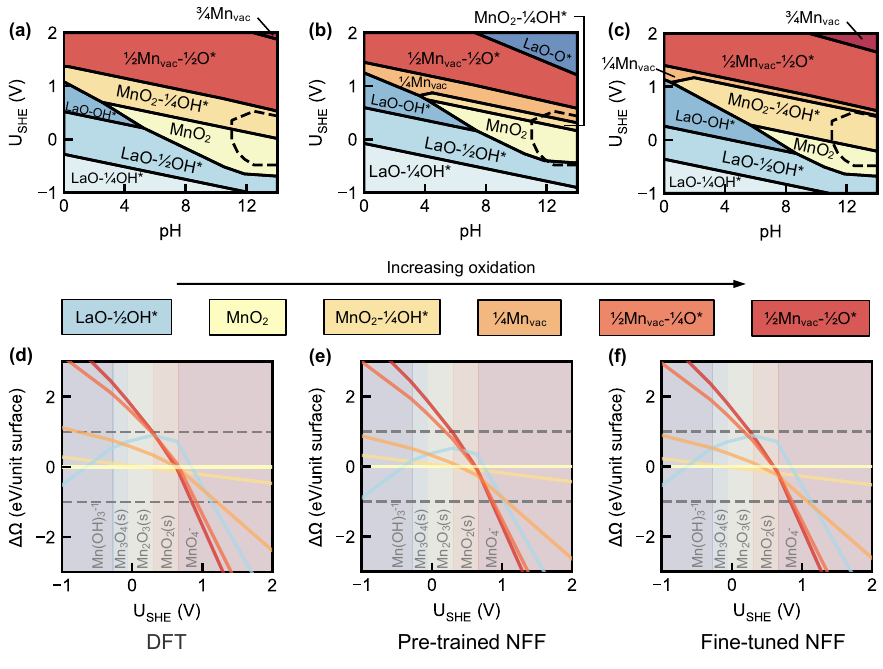}
  \caption{\textbf{\lmo{} surface Pourbaix diagrams and energy analysis with various energy models.} (a-c) \lmo{} surface Pourbaix diagrams from energies computed with (a) DFT, (b) pre-trained NFF, and (c) fine-tuned NFF. The \ce{LaMnO3} bulk stability regions at $10^{-6}$ M dissolved species concentrations are enclosed by the dashed lines. (d-e) $\Delta \grandpot$ comparison of handpicked \lmo{} surfaces with respect to the pristine surface in eV/surface unit cell across $\ushe$ at fixed pH = 12 with (d) DFT, (e) pre-trained NFF, and (f) fine-tuned NFF energies. The color sequence approximately follows increasing oxidation level. The stable species domains are differentiated through the background color, with \ce{La}\textsuperscript{+3} as the dominant La species and the dominant Mn species labeled in gray. Dotted lines are a guide for the eye.}
\label{fig:lmo}
\end{figure}

\subsubsection{\lmo}
Following our benchmark study on \pt{}, we evaluated our method on the more complicated \lmo{}. \ce{LaMnO3} is a perovskite with promising applications in both oxygen-evolution (OER) and oxygen-reduction (ORR) reactions \cite{rong_ab_2015, wan_enhanced_2025}. We evaluated 47 handpicked surfaces of the $2\times 2$ \lmo{} supercell from literature using DFT. These surfaces consist of varying levels of metal vacancies, and O* and OH* coverage levels. See Fig.~S5(a-b) for the pristine surface and Fig.~S5(c) for specific structures. A surface Pourbaix diagram for \lmo{} was constructed from these DFT calculations (Fig.~\ref{fig:lmo}(a))\cite{rong_ab_2015}. The \ce{LaMnO3} bulk phase stability region was obtained by mapping the corresponding solid entry in the \ce{LaMnO3} species Pourbaix diagram at $10^{-6}$ M for all dissolved species (Fig.~S10(b)). The general trend of the surface Pourbaix diagram agrees with a more oxidizing environment at higher $\ushe$ and pH, aligning with the species Pourbaix diagram and the analysis in \citet{rong_ab_2015}. We denote the surface structure by the stoichiometry of the surface termination, the fractional coverage of the prevailing adsorbate across sites, and the adsorbate species. For example, an \ce{MnO2} termination with $\frac{1}{4}$ of all metal sites covered with OH* is denoted \ce{MnO2}-$\frac{1}{4}$OH*, with a hyphen used for clarity. Meanwhile, vacancies and substitutions in the pristine surface layer are denoted by `vac' and `sub' subscripts respectively. At oxidizing conditions, surfaces are predominantly terminated by MnO\textsubscript{2}- or Mn\textsubscript{vac}-; in more reducing conditions at lower $\ushe$, LaO-type terminations become more preferred. Within each termination type, O* predominantly adsorbs at high $\ushe$ and pH, while OH* adsorption (protonation of O*) becomes prevalent under more reducing conditions. 

The surface energies of the handpicked structures were then evaluated using pre-trained CHGNet and the corresponding surface Pourbaix diagram plotted in Fig.~\ref{fig:lmo}(b). Notably, pre-trained CHGNet energies were already sufficient to recover most surface phases; while the predicted domain boundaries are not exact, they are well-aligned with those of the DFT surface Pourbaix diagram. The key differences are the replacement of $\frac{3}{4}$Mn\textsubscript{vac} in the top right (high $\ushe$ and high pH) of the DFT diagram with LaO-O* in the pre-trained CHGNet diagram, along with the emergence of an additional $\frac{1}{4}$Mn\textsubscript{vac} region in Fig.~\ref{fig:lmo}(b).

VSSR-MC sampling was again performed starting from the pristine surface with algorithmically generated virtual sites (see ``VSSR-MC'' in Methods for details) by sweeping across $\ushe$ and the pH range of 10 to 14, focusing on the bulk stability region, allowing for the adsorption and desorption of La, Mn, O, and OH (see Fig.~S6 for one such sampling profile). 

We downselected 118 VSSR-MC sampled surfaces and rattled the dominant structures found in the pre-trained CHGNet surface Pourbaix diagram to obtain another 18 structures (details in Methods ``Active learning''), resulting in a total of 136 structures for fine-tuning. Since pre-trained CHGNet predicted the surface Pourbaix diagram fairly well (Fig.~\ref{fig:lmo}(b)), we only fine-tuned CHGNet for \lmo{}. After fine-tuning, we re-evaluated energies for the handpicked structures and re-plotted the surface Pourbaix diagram in Fig.~\ref{fig:lmo}(c). This Pourbaix diagram has all but one phase agreeing with (the exception being a thin strip of $\frac{1}{4}$Mn\textsubscript{vac} present in Fig.~\ref{fig:lmo}(c)) and phase boundaries more aligned with the DFT reference, especially closer to the sampled bulk stability region with which we are concerned.

We plot $\Delta \grandpot$ with respect to the pristine \ce{MnO2}-terminated surface against $\ushe$ for the dominant surface phases by taking a slice along pH = 12 (Fig.~\ref{fig:lmo}(d-f)). As reflected in the surface Pourbaix diagrams, we find that most energies for the pre-trained CHGNet (Fig.~\ref{fig:lmo}(e)) are well-aligned with the DFT reference for \lmo{} (Fig.~\ref{fig:lmo}(d)) with a 0.095 eV/atom energy MAE (Fig.~S7(a)). This behavior contrasts with that of \pt{}, where pre-trained CHGNet was only able to predict qualitative trends with a much worse energy MAE of 0.531 eV/atom (Fig.~S4(a)). Only the energies of LaO-$\frac{1}{2}$OH* with respect to the pristine \ce{MnO2} and $\frac{1}{4}$Mn\textsubscript{vac} with respect to \ce{MnO2}-$\frac{1}{4}$OH* are underestimated. The improved pre-trained CHGNet performance could be attributed to the bulk structure of \ce{LaMnO3} already having oxygen atoms and thus the O*/OH* adsorbates on \lmo{} could be seen as less out of distribution compared with the thin O*/OH* layers on \pt{}.

After fine-tuning, we observe in Fig.~\ref{fig:lmo}(f) a large improvement in the relative surface Pourbaix grand potential of LaO-$\frac{1}{2}$OH* (from 0.379 eV/surface unit cell to 0.735 eV/surface unit cell at $\ushe = 0$) and a slight improvement in the relative surface Pourbaix grand potential of $\frac{1}{4}$Mn\textsubscript{vac} (from 0.314 eV/surface unit cell to 0.332 eV/surface unit cell at $\ushe =0$) to be closer to the DFT values in Fig.~\ref{fig:lmo}(d). We highlight that small changes of 10's or 100's of meV/surface unit cell is sufficient to change the relative ordering of surface stability, as in $\frac{1}{4}$Mn\textsubscript{vac}, and thus alter the resultant surface Pourbaix diagram; nevertheless, our fine-tuned NFF can very closely match the relatively-complicated DFT reference surface Pourbaix diagram for \lmo{}.


\begin{figure}[ht!]
  \centering
  \includegraphics[width=1.0\textwidth]{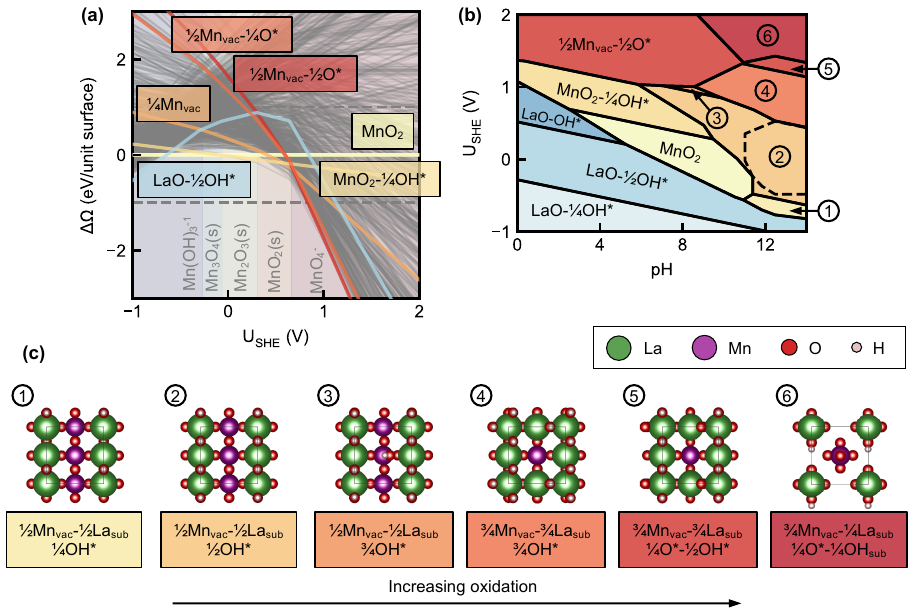}
  \caption{\textbf{\lmo{} energy analysis and surface Pourbaix diagrams.} (a) $\Delta \grandpot$ with respect to the pristine surface in eV/surface unit cell of handpicked and VSSR-MC sampled structures evaluated with fine-tuned CHGNet energies across $\ushe$ at fixed pH = 12. Handpicked structures are labeled while sampled structures are grayed. (b) Revised \lmo{} surface Pourbaix diagram at the DFT level with additional stable phases sampled using VSSR-MC. (c) Top-down view of stable sampled phases colored approximately in increasing oxidation level.}
\label{fig:sampled_structures}
\end{figure}

\subsection{Sampling additional surface phases}
Apart from sampling known reconstructions, our study of \lmo{} uncovered surfaces not identified in previous work. These newly sampled structures are represented as gray lines in the relative surface Pourbaix grand potential plot in Fig.~\ref{fig:sampled_structures}(a) using fine-tuned CHGNet energies (see Method ``Surface analysis'' for details) at pH = 12. Some of these additional surfaces are predicted to be even more stable than the low-energy handpicked structures. Subsequent DFT energy evaluations confirm their stability, leading to a revised \lmo{} surface Pourbaix diagram in Fig.~\ref{fig:sampled_structures}(b). In contrast, the sampled structures for \pt{} were not found to be lower in energy than known terminations, suggesting that existing terminations are already comprehensive. As a result, \pt{} surfaces were not further analyzed. (See Fig.~S8 for the equivalent \pt{} surface Pourbaix grand potential plot.)

The newly identified phases for \lmo{} (Fig.~\ref{fig:sampled_structures}(c)) demonstrate complex surface chemistry, featuring mixed La and Mn terminations that were absent from the handpicked structures but have been experimentally observed in \lmo{} reconstructions under ORR settings \cite{ignatans_effect_2019}. In Ref.~\citenum{ignatans_effect_2019}, enhanced ORR activity in \ce{LaMnO3} was attributed to the coexistence of \ce{Mn^{2+}} and \ce{Mn^{3+}} species at the surface layers. This mixed Mn oxidation state arises from the La and Mn mixed-termination that is distinct from the bulk structure. Among the VSSR-MC sampled structures, we identified a total of 23 such mixed-termination surfaces (Fig.~S9) more stable than the most stable surface among the handpicked structures (\ce{MnO2}-$\frac{1}{4}$OH*) at pH = 12 and $\ushe$ = 0.6 V. 

The stability trend of the new \lmo{} surface Pourbaix diagram (Fig.~\ref{fig:sampled_structures}(b)) with respect to electrical potential is consistent with our previous analysis of Fig.~\ref{fig:lmo}(a). At the most negative $\ushe$ at pH = 12, the most stable surface composition features a LaO termination with $\frac{1}{2}$OH*. Increasing $\ushe$ induces progressive oxidation, first by introducing Mn to form a mixed-termination structure, followed by a reduction of Mn occupancy in the termination layer. At the same time, more oxidizing species appear on the adlayer, first with OH* adsorption, followed by O* adsorption.

In summary, VSSR-MC sampled across compositions and configurations to discover over 1000 unique \lmo{} surface structures, far greater than the 47 handpicked surfaces initially considered, of which 6 structures were verified with DFT to be more stable than the handpicked structures.

\begin{figure}[ht!]
  \centering
  \includegraphics[width=1.0\textwidth]{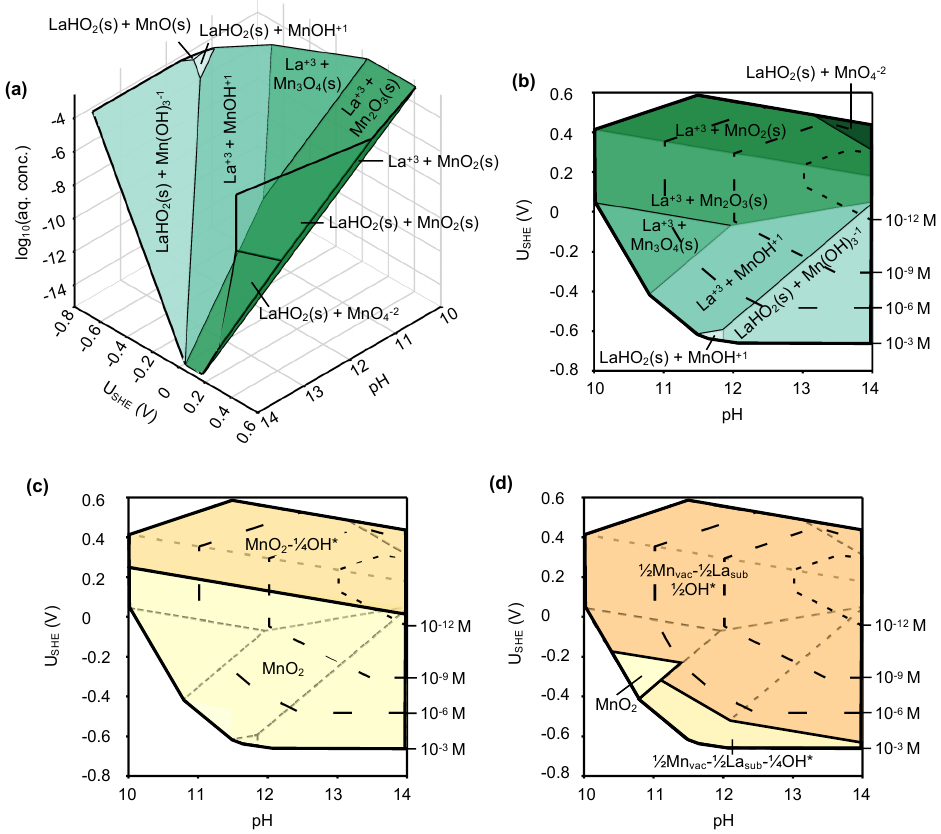}
  \caption{\textbf{\ce{LaMnO3} species and \lmo{} surface Pourbaix diagrams at thermodynamic equilibrium in the pH-$\ushe$-$\logconc$ axes.} Phases are colored approximately in increasing oxidation level from pale to dark. The black dashed contour lines correspond to different $\conc$. (a-b) Equilibrium species Pourbaix diagrams labeled by species domains adjacent to the bulk stability region. (a) 3D perspective. (b) 2D perspective in the pH-$\ushe$ axes viewed from high $\conc$. (c-d) 2D perspectives of equilibrium surface Pourbaix diagrams with (c) only literature surfaces and (d) including additional sampled surfaces.}
\label{fig:equilibrium}
\end{figure}

\subsection{New surface-environment equilibria conditions} \label{sec:new-eq}
After benchmarking our sampling method and NFF fine-tuning with conventional Pourbaix diagrams, we re-examine the thermodynamic assumptions and re-plot Pourbaix diagrams under bulk-electrolyte equilibrium conditions. We assumed a $10^{-6}$ M concentration for the stable dissolved species at each pH and $\ushe$ to generate the previous \lmo{} surface Pourbaix diagrams in keeping with literature convention. Here, we consider surface Pourbaix diagrams where $\conc$ is thermodynamically coupled to pH and $\ushe$ to account for equilibrium between the \ce{LaMnO3} bulk and metal species in the electrolyte. These equilibrium surface Pourbaix diagrams are a subset of the full 3D surface Pourbaix diagrams in pH-$\ushe$-$\conc$ space where $\conc$ varies as an independent variable alongside pH and $\ushe$.

Meanwhile, 3D surface Pourbaix diagrams were generated from the 3D \ce{LaMnO3} species Pourbaix diagram. Hence, as a first step towards generating the 3D surface Pourbaix diagrams, we extend the conventional \ce{LaMnO3} species Pourbaix diagram into 3D (Fig.~S11), where the dissolved species concentration varies as an independent variable alongside pH and $\ushe$, to compute thermodynamic equilibration between the electrode bulk and electrolyte species in the pH-$\ushe$-$\conc$ space. The resultant equilibrium species Pourbaix diagrams (Fig.~\ref{fig:equilibrium}(a-b)) show the \ce{LaMnO3} bulk-electrolyte species equilibrium Pourbaix region delineated from the 3D species Pourbaix diagram. (See ``Thermodynamic equilibria'' in Methods for details.) The domains are labeled according to the species adjacent to the bulk stability region in the 3D species Pourbaix diagram, which represent species present in the electrolyte. Fig.~\ref{fig:equilibrium}(a) shows the 3D perspective while Fig.~\ref{fig:equilibrium}(b) shows a projected view in the pH-$\ushe$ axes, with $\conc$ implicitly defined by thermodynamic equilibrium. The pH and $\ushe$ ranges spanned roughly correspond to the ORR region bounded by $-0.709 \text{ eV} < \ushe < 0.521 \text{ eV}$ at pH = 12, and are therefore much narrower than those in a conventional Pourbaix diagram. As we decrease $\conc$ from $10^{-3}$ M to $10^{-15}$ M, the stability region further narrows, aligning with Le Chatelier's principle on chemical equilibrium between the \ce{LaMnO3} bulk and adjacent species. From another perspective, any set of pH and $\ushe$ conditions within the bulk stability region corresponds to a pre-defined $\conc$. In principle, we can indefinitely extend the bulk stability region by increasing $\conc$. In practice, however, $\conc$ is limited by species solubility and device durability considerations \cite{kan_accelerated_2024}. 

Outside of the bulk stability region, \ce{LaMnO3} preferentially transforms into other La- and Mn-containing phases or dissolves entirely. In such cases, the assumption of a stable \ce{LaMnO3} bulk supporting the surface is no longer valid, as the underlying bulk structure itself changes. To ensure a meaningful surface Pourbaix analysis, we must therefore account for surface equilibration with both metal species in the electrolyte as well as the \ce{LaMnO3} bulk. We extend the conventional surface Pourbaix diagrams into 3D surface Pourbaix diagrams with an additional, independent $\conc$ axis in Figs.~S12 \& S13, and extract the same equilibrium Pourbaix region to respectively generate the equilibrium surface Pourbaix diagram with literature reported surfaces in Fig.~\ref{fig:equilibrium}(c) and the equilibrium surface Pourbaix diagram with both literature and additional sampled structures in Fig.~\ref{fig:equilibrium}(d). Like Fig.~\ref{fig:equilibrium}(b), these plots are 2D projections that resemble conventional Pourbaix diagrams across pH-$\ushe$ axes but with $\conc$ implicitly defined. Full 3D views in Fig.~S14 also illustrate this thermodynamic coupling between pH, $\ushe$, and $\conc$. The pH and $\ushe$ ranges are constrained by the equilibrium species Pourbaix diagram in Fig.~\ref{fig:equilibrium}(a-b), resulting in a narrower range compared to a conventional surface Pourbaix diagram, which shares limits with the conventional species Pourbaix diagram. Consequently, Fig.~\ref{fig:equilibrium}(c-d) show fewer phases than Figs.~S12 \& S13. Fig.~\ref{fig:equilibrium}(c) contains exactly two phases---pristine \ce{MnO2} and \ce{MnO2}-$\frac{1}{4}$OH*. With the inclusion of sampled surfaces in Fig.~\ref{fig:equilibrium}(d), the \ce{MnO2}-$\frac{1}{4}$OH* phase disappears and the \ce{MnO2} stability region shrinks. Most of the surface Pourbaix diagram is now covered by two newly-identified surfaces in Fig.~\ref{fig:sampled_structures}(c), with $\frac{1}{2}$Mn\textsubscript{vac}-$\frac{1}{2}$La\textsubscript{sub}-$\frac{1}{4}$OH* occupying an elongated region at the bottom while $\frac{1}{2}$Mn\textsubscript{vac}-$\frac{1}{2}$La\textsubscript{sub}-$\frac{1}{2}$OH* dominating the remainder. Other sampled surfaces from Fig.~\ref{fig:sampled_structures}(c) do not appear here as they fall outside of the bulk stability region. The slightly-warped phase boundary of the $\frac{1}{2}$Mn\textsubscript{vac}-$\frac{1}{2}$La\textsubscript{sub}-$\frac{1}{4}$OH* phase arises due to variations in $\conc$. 

In an aqueous electrochemical system where the electrode equilibrates with the electrolyte, the equilibrium surface Pourbaix diagram illustrates how the dominant surface phase evolves as a function of pH, $\ushe$, and $\conc$. Any two of these three variables uniquely determine the third; for instance, fixing $\conc$ and $\ushe$ allows us to determine the equilibrium pH in addition to the predominant surface. These diagrams can be leveraged to optimize catalyst reaction conditions by precisely targeting high-activity yet stable phases. Alternatively, they can inform the necessary conditions for forming a desired surface structure, aiding in the design of synthesis conditions. By capturing the interplay between pH, electrical potential, and the actual concentration of dissolved ions, these equilibrium surface Pourbaix diagrams provide a more comprehensive and useful picture of \lmo{} interface stability.

\section{Discussion}
We adapted VSSR-MC to tackle the challenging case of multi-component solid-liquid interfaces under electrical potentials and showed a pre-trained machine learning force field can describe trends in surface Pourbaix grand potentials. By sampling surface reconstructions with a pre-trained NFF and later fine-tuning with the generated structures, we can accurately predict surface Pourbaix diagrams and elucidate additional phases beyond those selected by human experts. Our new thermodynamic framework also respects equilibration across electrolyte species, the electrode surface, and the electrode bulk, providing a more faithful picture of aqueous electrochemical stability. 

In this work, we employed the computational hydrogen electrode framework to study surface stability, but its simplified assumptions do not fully capture the solvent-mediated electrostatic interactions at the electrode–electrolyte interface \cite{peng_toward_2025}. Specifically, the CHE model assumes that the electrochemical potential of aqueous species is governed solely by chemical activity, effectively setting the electrolyte as the reference for zero electrostatic potential. In reality, both the electrolyte and electrode are likely charged, forming an electric double layer that influences surface relaxation and stability predictions. Additionally, recent work \cite{yoon_quantum_2025} highlights how shifts in Fermi energy inside the working electrode correspond to shifts in the Galvani potential, introducing surface charge effects that influence both surface thermodynamics and reaction kinetics. Incorporating solvation models and grand canonical DFT \cite{hormann_grand_2019} into VSSR-MC could help capture these electrostatic contributions more accurately in future studies.

We also acknowledge limitations stemming from the finite simulation cell size and our focus on a single layer of virtual sites defined above the pristine surface. As a result, our approach may overlook more complex reconstructions involving multiple virtual site layers or deeper bulk rearrangements. However, given that NFFs have been shown to generalize well to larger supercells \cite{winter_simulations_2023, millan_effect_2023, roy_learning_2024}, expanding the configurational space in future studies should enhance our understanding of emergent phenomena under operational conditions.

Additionally, our current framework does not explicitly account for free energy contributions from vibrational or configurational entropy of the surface slab, which could play a significant role in determining phase stability at finite temperatures \cite{reuter_composition_2001, peng_data-driven_2024, peng_toward_2025}. Incorporating harmonic approximations for vibrational corrections and exploring configurational entropy effects in future extensions of VSSR-MC would improve the accuracy of surface free energy predictions and better capture entropic stabilization effects.

Finally, in our surface-environment equilibria analysis, we assumed equilibrium of the aqueous dissolved species with the electrode bulk without passivation effects that may occur at the interface. We also assumed all dissolved species are of the same concentration. Additional computational and experimental studies, especially within the ORR region (roughly the bulk stability region of \ce{LaMnO3}) \cite{rong_ab_2015, wan_enhanced_2025}, can help us determine the physical validity of our structures and their impact in modulating ORR reactions.

\section{Methods}
\subsection{VSSR-MC} \label{sec:methods-vssr-mc}
VSSR-MC\cite{du_machine-learning-accelerated_2023}  was performed in the semigrand-canonical ensemble with algorithmically generated virtual surface sites. The pH and $\ushe$ were supplied to the simulation and the total number of adsorbates may vary across an MC run. For each semigrand VSSR-MC iteration, one adsorption site was randomly chosen to change state.

For both \pt{} and \lmo, the sampling temperature was fixed at 3,000 K. For \pt, VSSR-MC was run at various $\ushe \in [0.8, 1.2]$ V in intervals of $0.1$ V and pH $\in [0, 12]$ in intervals of $4$. For \lmo, VSSR-MC was run at various $\ushe \in [-1.0, 2.0]$ V in intervals of $0.5$ V and pH $\in [10, 14]$ in intervals of $2$. The number of sweeps was fixed at 100 with 20 iterations each for a total of 2000 iterations for each $\ushe$ and pH value. See Figs.~S3 \& S6 respectively for \pt{} and \lmo{} sampling profiles.

Following the discrete sampling step, continuous relaxation was performed using the FIRE algorithm in the ASE package \cite{larsen_atomic_2017}. The convergence criterion was met when either a maximum of 20 relaxation steps was reached or the maximum force on all atoms fell below 0.05 eV/Å.

For each iteration in an MCMC run, the acceptance probability $P$ is given by the minimum of unity and the ratio of the Boltzmann weights between the proposed and current state:
\begin{align}
    P &= \min{\left\{1, \exp\left(-\frac{\Delta \grandpot}{k_\text{B} T_{\text{sample}}}\right)\right\}}
\end{align}
where $\Delta \grandpot$ is the change in surface Pourbaix grand potential after both discrete and continuous sampling (see ``Surface Pourbaix grand potential" in Methods for details), $k_\text{B}$ is the Boltzmann constant, and $T_{\text{sample}}$ is the sampling temperature. 

        
\subsection{Surface slab modeling}
\subsubsection{\pt{}}
A \ce{Pt} cubic unit cell from the Materials Project (mp-126) was cut in the (111) plane to create $2 \times 2$, $3 \times 3$, and $\sqrt{3}\times3$ supercells each with four Pt layers. A vacuum spacing of 15 Å in total was set at the ends of each slab. The bottom two layers were fixed while all other atoms were allowed to relax. A total of 24 and 54 virtual adsorption sites were respectively defined for the $2 \times 2$ and $3 \times 3$ supercells at top, bridge, and hollow sites. For the $\sqrt{3}\times3$ supercell, 6 virtual adsorption sites were defined at the top site. All sites sat at the default distance of 2.0 Å from the pristine surface using the pymatgen \verb|AdsorbateSiteFinder| class. Symmetry reduction of sites was disabled. Additionally, the atomic positions of the pristine surface were included in the virtual sites. 

The respective Pt virtual sites can be found in Fig.~S2(a-b). For the $2 \times 2$ and $3 \times 3$ structures, Pt, \ce{O}, and \ce{OH} were possible adsorbates, while for $\sqrt{3} \times 3$ structures, Pt, \ce{O}, \ce{OH}, and \ce{H2O} were possible adsorbates. The 8 handpicked surfaces were extracted from Hansen \textit{et al.}\cite{hansen_surface_2008} and Vinogradova \textit{et al.}\cite{vinogradova_quantifying_2018}. Other than the pristine surface, the specific composition of each handpicked slab is documented in Fig.~S2(c). These handpicked surfaces were optimized using DFT (see ``Density-functional theory calculations'' in Methods).

\subsubsection{\lmo{}}
A \ce{LaMnO3} cubic unit cell from the Materials Project (mp-19025) was cut in the (001) plane to create a $2 \times 2$ supercell with three bilayers of \ce{MnO2} and LaO, 6 layers in total. A total vacuum spacing of 15 Å was set at the ends of the slab. The bottom three layers were fixed while all other atoms were allowed to relax. A total of 24 virtual adsorption sites were defined at all top sites at the default distance of 2.0 Å from the \ce{MnO2} pristine surface using the pymatgen \verb|AdsorbateSiteFinder| class. Symmetry reduction of sites was disabled. Additionally, the atomic positions of the pristine surface were included in the virtual sites.  

All the virtual sites can be found in Fig.~S5(a-b). For this surface, La, Mn, O, and OH were possible adsorbates. The 47 handpicked surfaces were extracted from \citet{rong_ab_2015} and modified to be asymmetric as in our choice of slab. The specific composition of each handpicked slab is documented in Fig.~S5(c) and on Zenodo \cite{du_2025_data_accelerating}. These handpicked surfaces were also optimized using DFT (see ``Density-functional theory calculations'' in Methods).


\subsection{Species Pourbaix diagram construction}
The species Pourbaix diagrams for \ce{Pt} and \ce{LaMnO3} were calculated using a combination of data retrieval and processing steps \cite{persson_prediction_2012, singh_electrochemical_2017, patel_efficient_2019}. For each material, the combined metal (non-O, non-H) species Pourbaix diagrams were constructed using data obtained from the Materials Project database with the \texttt{MPRester} API and processed using both \\ \texttt{MaterialsProject2020Compatibility} and \texttt{MaterialsProjectAqueousCompatibility} compatibility schemes in \texttt{pymatgen} to ensure data consistency. The free energies of formation for stable solid entries were calculated and integrated with ion reference data to create species Pourbaix diagrams for pH $\in [0, 14]$ and $\ushe \in [-1.0 \text{ V}, 2.0 \text{ V}]$. Fig.~S10 shows species Pourbaix diagrams of \ce{Pt} and \ce{LaMnO3}. These species Pourbaix diagrams contain multiple domains with a unique set of stable species. During sampling, the \ce{LaMnO3} bulk region was replaced with individual La and Mn species to allow single-atom dissolution or re-deposition MC steps (Fig.~S10(b)). For evaluation of thermodynamic equilibria in pH-$\ushe$-$\conc$, all phases were considered (Fig.~S11). See ``Thermodynamic equilibria'' in Methods for details.

\subsection{Surface Pourbaix grand potential}\label{methods:pbx_formulation}
We reformulated the grand potential used in VSSR-MC by adding pH and $U$ dependencies: $\Omega_\text{surf} (\mathbf{X}, \{\mu\}, T, \ushe, \textpH)$. A grand potential formulation for bulk materials under aqueous electrochemical conditions was previously proposed by Refs.~\citenum{singh_electrochemical_2017,sun_non-equilibrium_2019, wang_optimal_2024} but here our focus is on surfaces formed from the dominant species at each pH and $\ushe$. The grand potential of any slab was calculated in two steps, inspired by the adsorption-dissolution framework described in Refs.~\citenum{rong_ab_2015, qiu_ab_2018} and generalizes the single-atom dissolution process in Fig.~\ref{fig:method} (a) to compute an absolute energy.
\begin{enumerate}
  \item All atoms were dissociated from the surface slab to their standard states, i.e., bulk for metal atoms, \ce{H2(g)} for hydrogen, and \ce{O2(g)} for oxygen. This energy change is computed here as: 
  \begin{align}
    \Omega_1 = \sum_{\elemA \in \mathcal{A}} N_\elemA \muA\stst - G_{\text{slab}}
  \end{align}
  where $G_{\text{slab}}$ is the Gibbs free energy of the slab, $\mathcal{A}$ is the set of elements present in the slab, element $N_\elemA$ is the number of atoms of element $A$, $\mu_A \stst$ is the standard state chemical potential of element $
  \elemA$. $G_{\text{slab}}$ was estimated with the slab energy calculated using the prescribed energy model, e.g., DFT or NFF. Raw DFT energies were corrected with \texttt{MaterialsProject2020Compatibility}. Additionally, a zero-point energy - $TS$ (ZPE-TS) correction of $0.23$ eV was added to each OH adsorbate \cite{rong_ab_2015} on the slab. $\mu_\elemA \stst$ values were obtained with \texttt{MPRester} and then corrected with \texttt{pymatgen}. 
  
  \item Each standard state species $\elemA$ was reacted to form the most stable species under aqueous electrochemical conditions with the general formula \ce{H_x A O_y^{z-}}, where x, y, and z are integer coefficients. The chemical equation for each individual A is as follows:
  \begin{align}\label{eqn:std_state_to_aq_eq}
    \ce{A} + N_{\ce{A}, \ce{H_2O}} \ce{H_2O -> H_x AO_y^{z-}} + N_{\ce{A},\ce{H^+}} \ce{H+} + N_{\ce{A},\ce{e}} \ce{e-}
  \end{align}
  where $N_{\ce{A}, \ce{H_2O}}$, $N_{\ce{A},\ce{H^+}}$, and $N_{\ce{A},\ce{e}}$ are the stoichiometric coefficients of \ce{H2O}, \ce{H+}, and \ce{e-} in the reaction, respectively.
  The dominant aqueous electrochemical species A depends on the pH, $\ushe$, $T$, and $\muspecies$. For this work, we calculated energies at $T = 298$ K and $\muspecies = \mu_{\mathrm{H_x A O_y^{z-}}}\stst + \kb T \ln a_\mathrm{H_x A O_y^{z-}}$, where $\mu_{\mathrm{H_x A O_y^{z-}}}\stst$ is the standard state energy of the species and $a_\mathrm{H_x A O_y^{z-}}$ is the species activity, which we assume to be 1 for solid species and $10^{-6}$ for dissolved species unless specified. The most stable species was queried at each set of pH and $\ushe$, which effectively is the dominant phase present in the species Pourbaix diagrams (see previous section). The total energy change for step 2 is as follows: 
  \begin{align}\label{eqn:std_state_to_aq_energy}
    \Omega_2 &= \sum_{A \in \mathcal{A}} \muspecies - \muA -  N_{\ce{A}, \ce{H_2O}} \mu_{\ce{H_2O}} - 2.3 N_{\ce{A},\ce{H^+}}  \kb T \text{pH} - N_{\text{A}, \text{e}} (e \ushe)  \nonumber \\
    &= \sum_{A \in \mathcal{A}} \Delta \Omega_{\text{A, SHE}} \stst - 2.3 N_{\ce{A},\ce{H^+}}  \kb T \text{pH} - N_{\text{A}, \text{e}} (e \ushe) + \kb T \ln a_\mathrm{H_x A O_y^{z-}} 
  \end{align}
  where $\Delta \Omega_{\text{A, SHE}} \stst$ is the standard-state free energy of reaction for species A under SHE and is in accordance with the CHE formulation.
\end{enumerate}

The surface Pourbaix grand potential is defined as the formation energy from the stable aqueous species, $\Omega_\text{surf} (\mathbf{X}, \{\mu\}, T, \ushe, \textpH) = -(\Omega_1 + \Omega_2$), with the negative sign.

\subsection{Surface analysis}
\subsubsection{\pt{}}
Following each VSSR-MC run, the sampled surfaces were initially downselected to 100 structures from the original 2000 generated per run. $2\times 2$ and $3\times 3$ structures with more than 4 and 9 O atoms, respectively, were filtered out to mitigate excessive \ce{O2} adsorption introduced by \texttt{MaterialsProject2020Compatibility} corrections, yielding 1131 and 979 structures, respectively. All 2000 $\sqrt{3}\times 3$ structures were initially included. After removing duplicate \pt{} pristine surfaces, a total of 3964 unique surfaces, along with the 8 handpicked structures, were evaluated using fine-tuned MACE to generate the relative surface Pourbaix grand potential plot in Fig.~S8.

\subsubsection{\lmo{}}
Following each VSSR-MC run, the sampled surfaces were also downselected to 100 structures from the original 2000 generated per run. Structures with 8 or fewer Mn atoms but more than 38 O atoms were filtered out to mitigate excessive \ce{O2} adsorption caused by \texttt{MaterialsProject2020Compatibility} corrections, yielding 1319 structures. An additional 7 structures were removed due to unphysical bonding configurations, such as floating O* or OH* on oxygen atoms. The remaining 1312 surfaces, along with the 47 handpicked surfaces, were then evaluated using fine-tuned CHGNet to generate the relative surface Pourbaix grand potential plot in Fig.~\ref{fig:sampled_structures}(a).

Subsequently, 27 low-energy structures were selected based on their fine-tuned CHGNet $\grandpot$ values at $\ushe$ = 0.6 V and pH = 12, specifically those at most 0.1 eV/surface unit cell higher in energy than the initially most stable surface, \ce{MnO2}-$\frac{1}{4}$OH*. These structures were further relaxed with DFT, resulting in 23 surfaces that were more stable at the DFT level than \ce{MnO2}-$\frac{1}{4}$OH* under the same conditions (Fig.~S9). Integrating these structures with those from literature for convex hull analysis resulted in the updated surface Pourbaix diagram shown in Fig.~\ref{fig:sampled_structures}(b).

\subsection{Thermodynamic equilibria}
To extend the 2D species and surface Pourbaix diagrams from pH-$\ushe$ space into 3D pH-$\ushe$-$\conc$ space, we allowed the chemical potentials of dissolved species to vary with concentration. In particular, the species chemical potential, $\muspecies$, is related to its activity, $a_\mathrm{H_x A O_y^{z-}}$, through: $\muspecies = \mu_{\mathrm{H_x A O_y^{z-}}}\stst + \kb T \ln a_\mathrm{H_x A O_y^{z-}}$ and under dilute concentrations ($c_\mathrm{H_x A O_y^{z-}} \leq 10^{-3}$ M), the activity can be approximated by the concentration $a_\mathrm{H_x A O_y^{z-}} \approx c_\mathrm{H_x A O_y^{z-}}$. 

Using this relationship, we performed convex hull analyses over a four-dimensional (4D) space (pH-$\ushe$-$\conc$ and the energy) to construct 3D Pourbaix diagrams (Figs.~S11, S12 \& S13). First, we generated the 3D \emph{species} Pourbaix diagram with convex hull analysis to determine the dominant species A across pH-$\ushe$-$\conc$ space. Next, we expressed the \emph{surface} Pourbaix grand potential as a function of an additional variable, $\Omega_\text{surf} (\{c\}, \ushe, \textpH)$, where $\{c\}$ represents the set of concentrations for each dominant dissolved species, $c_\mathrm{H_x A O_y^{z-}}$. We do this analysis for each set of dominant species, A, as in Fig.~S1, and separately for both Figs.~S12 \& S13. A second round of convex hull analyses in 4D space yielded the 3D \emph{surface} Pourbaix diagrams.

These 3D Pourbaix diagrams have independent pH, $\ushe$, and $\conc$ axes and thus are not under full equilibration between the electrode bulk, electrode surface, and electrolyte. To achieve full equilibrium, each species neighboring the \ce{LaMnO3} bulk phase in the \emph{species} Pourbaix diagram must be in equilibrium with the bulk itself, i.e., at the phase boundaries where they meet. For example, consider \ce{La^{+3}} + \ce{MnO2(s)} (electrolyte) in equilibrium with \ce{LaMnO3(s)} (electrode) at $10^{-6}$ M dissolved species concentrations in Fig.~S10(b). The energy changes from standard-state species were evaluated using Eqns.~\eqref{eqn:std_state_to_aq_eq} \& \eqref{eqn:std_state_to_aq_energy} and set equal, leading to the equilibrium condition:
\begin{align}
    \mu_{\text{\ce{LaMnO3}}} +2 \cdot 2.3 \kb T \textpH &= \mu_{\text{\ce{La^{3+}}}} + \mu_{\text{\ce{MnO2}}} + \mu_{\text{\ce{H2O}}} + e \ushe
\end{align}

Here, $\mu_{\text{\ce{LaMnO3}}} \approx E_{\text{\ce{LaMnO3}}}^{\text{DFT}}$, $\mu_{\text{\ce{MnO2}}} \approx E_{\text{\ce{MnO2}}}^{\text{DFT}}$, and $\mu_{\text{\ce{H2O}}} \approx -2.46 \text{ eV/\ce{H2O}}$. This equation leaves three adjustable parameters (pH, $\ushe$, and $c_{\ce{La^{3+}}}$) but only two degrees of freedom. By applying analogous equilibrium conditions for every species adjacent to \ce{LaMnO3} in the 3D \emph{species} Pourbaix diagram (pH $\in [0, 14]$, $\ushe \in [-1.0 \text{ V}, 2.0 \text{ V}]$, and $\conc \in [10^{-15} \text{ M}, 10^{-3} \text{ M}]$), we effectively trace out the \ce{LaMnO3} bulk stability boundaries and label the neighboring species at each point, leading to the equilibrium \textit{species} Pourbaix diagram where pH, $\ushe$, and $\conc$ are thermodynamically coupled. In practice, we introduced a slightly offset hyperplane representing the \ce{LaMnO3} phase in the 4D space, oriented opposite to the original \ce{LaMnO3} plane. 

Finally, we traced out the same bulk stability boundaries from the 3D \emph{surface} Pourbaix diagrams in Figs.~S12 \& S13 to enforce full electrode-surface-electrolyte equilibrium and obtain the equilibrium \emph{surface} Pourbaix diagrams, with pH, $\ushe$, and $\conc$ also thermodynamically coupled. This approach parallels the well-established framework for thermodynamic equilibria in contact with gas \cite{reuter_composition_2001, heifets_density_2007, heifets_electronic_2007, todd_selectivity_2021, du_machine-learning-accelerated_2023, chen_geometry_2024}. However, unlike gas-solid interfaces where species identities are fixed, aqueous species depend on pH, $\ushe$, and $\conc$.

\subsection{Active learning}
For \pt{}, the medium MACE-MP-0 pre-trained model \cite{batatia_mace_2022} was fine-tuned for 60 epochs. The loss function was a weighted sum of the mean-squared errors of forces and energy with a 100:1 ratio for weights of forces and energies, respectively. The node embedding, pooling, and readout layers were fully unfrozen while for the interaction layers only the linear layers were allowed to changed. For \lmo{}, pre-trained CHGNet \cite{deng_chgnet_2023} v0.3.0 was fine-tuned for 60 epochs. The loss function was a weighted sum of the mean-squared errors of forces and energy with a 100:5 ratio of weights. For pre-trained CHGNet, unfreezing the readout layer was sufficient for good fine-tuning performance on \lmo{}. For both systems, raw DFT energies were corrected with the \texttt{MaterialsProject2020Compatibility} scheme and the Adam optimizer \cite{kingma_adam_2015} was used with a learning rate of 0.001.

Structures were selected for active learning using latent space clustering, as in \citet{du_machine-learning-accelerated_2023}. Briefly, the VSSR-MC generated structures for each surface were clustered according to the first three principal components (PCs; $>90\%$ explained variance) of their NFF embeddings and the most uncertain structure for each cluster was selected. To save on compute time, a first-pass clustering was run for every 1,000 samples to yield around 100-200 structures at each set of pH and $\ushe$. The results of one full clustering run for each of \pt{} and \lmo{} is shown in Fig.~S15. 

The metric for uncertainty quantification was changed to Gaussian-mixture model (GMM) uncertainty, which is more suitable for a single NFF model used here. \cite{tan_single-model_2023}. The GMM model was calibrated against the force MAE of a 5000-structure subset randomly selected from MPtrj \cite{deng_materials_2023}, which was used for training the foundational models. We show the uncertainty vs. force MAE plot in Fig.~S16 for the MPtrj calibration dataset and all sizes of \pt{} and \lmo{} fine-tuning structures. The performance is commensurate with previous work \cite{zhu_fast_2023, tan_single-model_2023}.

Additionally, \pt{} and \lmo{} structures in the pre-trained CHGNet surface Pourbaix diagrams were perturbed by randomly displacing atoms in all three spatial directions, with a maximum displacement of $\pm 0.1$ Å, to better sample the near-equilibrium potential energy surface. A single active learning cycle was sufficient to refine the respective NFF models, reducing their MAE to within 10 meV/atom relative to DFT references across all surfaces, as validated against handpicked literature structures (see Figs.~S4 \& S7).

\subsection{Density-functional theory calculations}
Vienna \textit{ab initio} Simulation Package (VASP) v.6.2.1 \cite{kresse_efficiency_1996, kresse_efficient_1996} was employed for both single-point DFT calculations and DFT relaxations using the projector augmented-wave (PAW) method to describe core electrons \cite{blochl_projector_1994, kresse_ultrasoft_1999}. The following PAW datasets version 54 pseudopotentials were used: H, O, Pt, La, and Mn\_pv. The Perdew-Burke-Ernzerhof (PBE) functional \cite{perdew_generalized_1996} version of the generalized-gradient approximation (GGA) was used for all  calculations. Grimme’s D3 method with Becke-Johnson damping was used to account for dispersion interactions \cite{grimme_consistent_2010, grimme_effect_2011}. All calculations were spin-polarized. Additionally, DFT+U calculations were used for \lmo{} applying Dudarev's approach \cite{dudarev_electron-energy-loss_1998} with the single on-site parameter for Mn set to 3.9 on the d-orbital electrons, consistent with the Materials Project settings \cite{jain_commentary_2013}. Dipole corrections to the total energy were enabled along the z-axis for surfaces. The kinetic energy cutoff for plane waves was set to 520 eV. Integrations over the Brillouin zone were performed using a $6 \times 6 \times 1$ Gamma centered mesh for $2\times2$ \pt{}, $4\times4\times1$ Gamma centered mesh for $3\times3$ \pt{}, $6\times4\times1$ Gamma centered mesh for $\sqrt{3}\times3$ \pt{}, and a $3\times3\times1$ Gamma centered mesh for \lmo{}. In the self-consistent field cycle, a total energy limit of 10\textsuperscript{-6} eV was adopted as the stopping criterion.

For surfaces that were optimized, optimization of atomic positions was performed until the Hellmann-Feynman forces on atoms were smaller than 10 meV/Å or until 100 relaxation steps were reached using the conjugate gradient algorithm for optimization and for electronic minimization within each self-consistent field calculation. After 100 relaxation steps, any unconverged surfaces were further relaxed with the RMM-DIIS algorithm until the change in total energy between two relaxation steps was smaller than 10 meV.



\subsection{Workflow management and compute time}\label{methods:workflow}
An internal library, HTVS (for high-throughput virtual simulations) managed the DFT calculations. VSSR-MC and active learning were run in separate procedures. DFT single-point calculations took about 15 minutes to 1 hour each on an NVIDIA Volta V100 32 GB GPU (MIT SuperCloud) while relaxations took about 3-6 hours each on an NVIDIA Tesla A100 Ampere 40 GB GPU (NERSC Perlmutter). VSSR-MC runs were $2000$ iterations each. For \pt{}, $2\times2$ slab runs took 30 minutes each, $3\times3$ slab runs took 50 minutes each, while $\sqrt{3}\times3$ slab runs took 40 minutes each on an NVIDIA GeForce RTX 2080 Ti 11 GB GPU. Meanwhile, $2\times2 $ \lmo{} runs took 70 minutes each. Active learning was relatively fast, with latent-space clustering and fine-tuning taking $15$ minutes or less on an RTX 2080 Ti 11 GB GPU.

\section{Data availability}
The fine-tuned models, DFT data, selected results from VSSR-MC runs, and Jupyter notebooks used for data analysis and plotting are available on Zenodo: \url{https://doi.org/10.5281/zenodo.15066440}.

\section{Code availability} 
The VSSR-MC algorithm reported in this work is available on GitHub: \url{https://github.com/learningmatter-mit/surface-sampling}. Our version of \texttt{pymatgen} for constructing surface Pourbaix and 3D Pourbaix diagrams, along with additional plotting methods, is also available on GitHub: \url{https://github.com/xiaochendu/pymatgen}.

\begin{acknowledgement}
The authors appreciate manuscript editing by Juno Nam. X.D. acknowledges support from the National Science Foundation Graduate Research Fellowship under Grant No. 2141064. The authors acknowledge the MIT SuperCloud and Lincoln Laboratory Supercomputing Center for providing HPC resources that have contributed to the research results reported within this paper. This research used resources of the National Energy Research Scientific Computing Center (NERSC), a Department of Energy Office of Science User Facility using NERSC awards BES-ERCAP-m4604 and BES-ERCAP-m4866.
\end{acknowledgement}

\begin{suppinfo}
A list of abbreviations and Supplementary Figures are available in the SI.
\end{suppinfo}

\providecommand{\latin}[1]{#1}
\makeatletter
\providecommand{\doi}
  {\begingroup\let\do\@makeother\dospecials
  \catcode`\{=1 \catcode`\}=2 \doi@aux}
\providecommand{\doi@aux}[1]{\endgroup\texttt{#1}}
\makeatother
\providecommand*\mcitethebibliography{\thebibliography}
\csname @ifundefined\endcsname{endmcitethebibliography}
  {\let\endmcitethebibliography\endthebibliography}{}

\end{document}